\documentclass[a4paper,11pt]{article}
\usepackage{jheppub} 

\usepackage{amsmath}
\usepackage{dsfont}
\usepackage{bm}

\title{Black holes and solitons in an extended Proca theory}

\author[a]{Eugeny Babichev,}
\author[a]{Christos Charmousis,}
\author[b]{and Mokhtar Hassaine}

\affiliation[a]{Laboratoire de Physique Th\'eorique, CNRS, Univ. Paris-Sud, \\ Universit\'e Paris-Saclay, 91405 Orsay, France}
\affiliation[b]{Instituto de Matem\'atica y F\'{\i}sica, Universidad de Talca,
Casilla 747, Talca, Chile.}
\emailAdd{eugeny.babichev@th.u-psud.fr}
\emailAdd{christos.charmousis@th.u-psud.fr}
\emailAdd{hassaine@inst-mat.utalca.cl}

\abstract{We study a massive vector tensor theory that acquires mass via a standard Proca term but also 
via a higher order term containing an explicit coupling to curvature. 
We find static  solutions that are asymptotically flat, adS or Lifshitz black holes. 
Since the higher order term regularizes the effect of the Proca mass term, generically solutions are asymptotically regular for arbitrary couplings. 
This is true in particular for asymptotically flat black holes.
For a particular coupling we find particle like solitons that have a regular and non trivial geometry everywhere. In all adS solutions 
the Proca mass term plays the role of an effective cosmological constant distinctly different from the bare cosmological constant.}

\begin{document} 
\maketitle
\flushbottom

\section{Introduction}

The unknown nature of dark energy and dark matter question the validity of General Relativity at very large distance scales. As a result, modified gravity theories have attracted  a lot of attention recently and an important effort is made by the community to understand different facets of these theories~\cite{Clifton:2011jh}, their stability, their cosmological implications, the existence and nature of compact objects etc. Most of the attention has been given to scalar tensor theories as they are the simplest of modified theories, and additionally, present a lot of characteristics one finds in more complicated theories~\cite{Charmousis:2014mia}. 
If instead of a scalar field one considers a vector degree of freedom, this provides another way to modify gravity. 
In this paper we will study black hole and soliton solutions of vector tensor theories. It is instructive to overview vector tensor theories as they have appeared in many different facets since the advent of General Relativity (GR). 

Vector tensor theories have been approached in many different ways and since a long time. A convenient starting point{\footnote{The list references are indicative and not exhaustive-the interested reader should consult within the references provided here for a more complete bibliography. We are however including here some of the older key references.}} is that of the Maxwell field coupled with GR, Einstein-Maxwell theory (EM). This theory describing the standard model photon's interactions with gravity is massless, has two propagating vector degrees of freedom (for the photon) and has $U(1)$ Abelian gauge symmetry. Additionally, the theory has electro-magnetic duality and is conformal{\footnote{By conformal we mean that the energy momentum tensor of the electromagnetic field is traceless in 4 dimensions}} in $4$ dimensions. As a result EM theory cannot drive slow roll inflation. Furthermore, it cannot seed observed astrophysical magnetic fields (of the size of the micro-Gauss) as the magnetic field is strongly diluted during inflation (see for example the review \cite{Kandus:2010nw}); one has to go beyond EM theory in order to pursue such and other effects, for example, related to the CMB \cite{Ade:2013nlj}. 

Introducing more complicated kinetic terms one gets vector tensor theories studied early on \cite{Will:1972zz}. 
In such theories gauge symmetry is absent and the vector has generically 4 polarisations, of which, the time component of the vector is always a ghost. To remedy this problem Jacobson and collaborators considered spontaneous breaking of Lorentz symmetry by fixing the norm of the vector and thus introducing Einstein-Aether theory~\cite{Jacobson:2008aj}. This theory is a nice prototype theory where one can study in detail the breaking of Lorentz invariance. 

Another interesting approach has been to try to keep $U(1)$ gauge symmetry while asking if the EM action can be generalised. Horndeski in a series of nice papers studied an extension of EM \cite{Horndeski:1977kz, Horndeski:1978dw} for which he demonstrated that there is a unique additional term which can be added to EM enjoying the following properties: Maxwell equations in flat spacetime, $U(1)$ gauge symmetry and second order field equations 
(at the same time the electromagnetic duality and conformal coupling is absent in this extension, in contrast to the Maxwell theory). This term involves the double dual curvature tensor which modifies EM theory in the presence of curvature. In other words one can still consider this theory as a theory describing photons coupled to gravity, albeit nonminimally (for constraints on the variation of the fine structure constant see \cite{Uzan:2002vq}). 
Moreover, Horndeski  showed that there is a Birkhoff theorem \cite{Horndeski:1978dw} for such a theory. The additional term, akin to a "Paul" term in Fab 4 \cite{Charmousis:2011ea} theory, introduces a certain strain for far away asymptotics in a similar way to a Proca mass term, which we will turn to in a moment. Stability and cosmological implications of Horndeski-Maxwell theory have been studied more recently \cite{Barrow:2012ay}. This theory was shown to be a direct consequence of a Kaluza-Klein cascade of Lovelock theory very early on \cite{Buchdahl:1979wi} (for black hole solutions see \cite{Charmousis:2008kc}).

Another course of action is that of adding a mass term to the EM theory; thus losing $U(1)$ gauge invariance and dealing with an Einstein-Proca theory with an additional longitudinal degree of freedom (to the two vectorial ones of EM). If the Proca field is assumed to be a photon with mass then there are stringent constraints{\footnote{If the Proca mass is associated to dark energy and, therefore to the size of the universe, it is still within the allowed photon mass constraint bounds}} \cite{Dolgov:1981hv}. Furthermore, Proca theories introduce unusual asymptotics, similar to a Horndeski Maxwell term, it is in fact well known that such theories are more akin to Lifshitz spacetimes with anisotropic space and time scaling. We will see explicitly this arising here in the later sections. But as we will see to remedy the effects of Proca mass one can  consider higher order terms in the same manner as Horndeski considered scalar tensor theories (see \cite{Allys:2015sht} and references within). In part{\footnote{Certain intrinsic vector interactions are not obtainable from this simple substitution \cite{Heisenberg:2014rta}}} one can start with a shift symmetric Horndeski theory and replace $\nabla_\mu \phi$ by the vector $A_\mu$ \cite{Heisenberg:2014rta}. One then deals with an extension of Horndeski theory or vector-tensor galileons as constructed in \cite{Deffayet:2010zh}. One of course loses $U(1)$ gauge symmetry and introduces in some form or another a mass term for the vector. Adding a mass term has stringent constraints if our field is still the photon (coupled to the standard model (!)). Nevertheless one can consider theories where an effective mass is created but only due to the presence of curvature. In other words one could still have Maxwell equations in flat spacetime. This may have interesting cosmological implications. 

When one is seeking black hole solutions with additional fields one is confronted with no hair theorems. It is known that Einstein-Proca theories for example do not admit hairy solutions as was discussed by Bekenstein, \cite{Bekenstein:1971hc}. We will see that the additional curvature-vector term modifies this conclusion providing hairy black holes (see also \cite{Chagoya:2016aar} \cite{Herdeiro:2016tmi}) for asymptotically flat spacetimes. 
Putting these considerations together we will consider the following action
\begin{eqnarray}
S[g,A]=\int\,\sqrt{-g}\,d^4x\Big[R-2\Lambda -\frac{1}{4}{\cal
F}^2-\frac{\mu^2}{2}A^2+\beta G_{\mu\nu}A^{\mu}A^{\nu}\Big].
\label{action}
\end{eqnarray}
This action is Einstein-Proca for $\beta=0$ while it is EM with a higher order term if $\mu=0$. Recently there was some activity finding hairy solutions for the above theory \cite{Chagoya:2016aar}, \cite{Minamitsuji:2016ydr}, \cite{Cisterna:2017umf}. 
For the case $\mu=0$ the case of a particular coupling $\beta=1/4$ was studied \cite{Chagoya:2016aar}. There it was postulated that only for this particular coupling there existed asymptotically flat solutions. We will see explicitly that this is not the case. Asymptotically flat black holes can be found for generic values of the coupling $\beta$. Using previous work on black holes for scalar tensor theories \cite{Babichev:2013cya} (see also \cite{Rinaldi:2012vy}), it was realized \cite{Minamitsuji:2016ydr}, that there existed a simple procedure to relate certain scalar tensor to vector tensor solutions of (\ref{action}). A set of such solutions was obtained for certain couplings. We will undertake here a full analysis of the above action for static and spherical symmetry. We will find previous solutions in our analysis, we will discard some which are non relevant mathematically and will discover numerous others. We will most importantly classify the possible solutions and show that one gets asymptotically flat solutions for $\mu=0$ and generic coupling $\beta$ while one gets asymptotic adS solutions in the presence of a Proca term. Furthermore, allowing for the Proca mass term we will find everywhere regular solitons with the Proca term playing the role of an effective negative cosmological constant (see also \cite{Brito:2015pxa}) and also, less surprisingly, Lifshitz black holes (see also \cite{Geng:2015kvs}). To construct these, as we will see,  it is important to allow for both Proca and galileon terms. 

In the next section we set up the theory in question and then show how the field equations boil down to two coupled field equations, one algebraic and one ODE with respect to two variables. In section 3 we solve for a particular coupling in all generality and find black hole but also soliton solutions. In section 4 we solve for arbitrary coupling but by setting to zero an integration constant. Asymptotically flat,  adS but also Lifshitz solutions are found. We round up our results and perspectives in the final section.  

\section{Theory and field equations}

The variation with respect to the metric of (\ref{action}) reads,
\begin{eqnarray}
\label{einstein} {\cal{E}}_{\mu\nu}:=G_{\mu\nu}+\Lambda
g_{\mu\nu}-\frac{1}{2}\Big[F_{\mu\sigma}F^{\,\sigma}_{\nu}-
\frac{1}{4}g_{\mu\nu}F_{\alpha\beta}F^{\alpha\beta}\Big]-\frac{\mu^2}{2}\left(A_{\mu}A_{\nu}-\frac{1}{2}g_{\mu\nu}A^2\right)-\beta
Z_{\mu\nu}
\end{eqnarray}
where
\begin{eqnarray}
Z_{\mu\nu} &=& \frac{1}{2} A^2 R_{\mu\nu} + \frac{1}{2} R\, A_{\mu}
A_\nu -2A^\alpha R_{\alpha (\mu} A_{\nu)} -\frac{1}{2}
\nabla_\mu\nabla_\nu A^2 + \nabla_\alpha \nabla_{(\mu} \big(A_{\nu)}
A^\alpha\big)\cr &&-\frac{1}{2} \Box (A_\mu A_\nu) + \frac{1}{2}
g_{\mu\nu} \big(G_{\alpha\beta} A^\alpha A^\beta + \Box A^2 -
\nabla_\alpha\nabla_\beta (A^\alpha A^\beta)\big)\
\end{eqnarray}
while the modified Proca  equation reads,
\begin{eqnarray}
\label{proca}
J^{\nu}:=\nabla_{\mu}( F^{\mu\nu})-\mu^2 A^{\nu}+2\beta
A_{\mu}G^{\mu\nu}=0.
\end{eqnarray}
Take static and spherically symmetric spacetime
\begin{eqnarray}
&& ds^2=-h(r)dt^2+\frac{dr^2}{f(r)}+r^2d\Omega^2_{2,\kappa},\qquad
A_{\mu}dx^{\mu}=a(r)dt+\chi(r)dr.\label{ansatz}
\end{eqnarray}
where $\kappa$ corresponds to the curvature of the base manifold
$\kappa=0,\pm 1$. Spherical symmetry corresponds to $\kappa=1$. As we will also study asymptotically adS and Lifshitz spacetimes we allow here also for a hyperbolic and planar base manifold, $\kappa=-1,0$ respectively. 
It is important to notice the presence of $\chi(r)$ which is no
longer a gauge term in the presence of a massive vector field.

The equation $J^{r}=0$ implies either that $\chi(r)$ is trivial or alternatively
the metric constraint,
\begin{eqnarray}
f(r)=\frac{h(r)\left(\mu^2
r^2+2\beta\kappa\right)}{2\beta\left(r\,h\right)^{\prime}}.
\label{relfh}
\end{eqnarray}
We choose the latter option. 
The ${\cal{E}}_{tr}=0$ equation is then immediately verified and hence the system is mathematically consistent. The ${\cal{E}}_{rr}=0$ equation gives the $\chi(r)$ field
\begin{eqnarray}
\label{chi}
\chi^2(r)&=&\frac{r\Big[(\frac{\mu^2}{2} r^2+\beta\kappa)\left(\beta a^2 h'-2\beta a a' h'-\frac{1}{4} r h (a')^2\right)-\frac{1}{2}(r^2h^2)'(\frac{\mu^2}{2}+\beta\Lambda)\Big]}{h^2(\frac{\mu^2}{2} r^2+\beta\kappa)^2}
\end{eqnarray}
At the end we are left with the other non trivial component of the Maxwell-Proca equation $J^{t}=0$ and the metric equation ${\cal{E}}_{tt}=0$. These latter two equations are simplified noting the substitution \cite{Babichev:2013cya}
\begin{eqnarray}
h(r)=-\frac{2M}{r}+\frac{1}{r}\int \frac{k(r)}{\mu^2 r^2+2\beta\kappa} dr
\label{susb}
\end{eqnarray}
yielding at the end,
\begin{equation}
\label{master1} \left[\frac{(\mu^2 r^2+2\beta\kappa)(r\;
a)'}{\sqrt{k(r)}}\right]'=(1-4\beta) a(r) \left[\frac{(\mu^2
r^2+2\beta\kappa)}{\sqrt{k(r)}}\right]'
\end{equation}
\begin{equation}
\label{master2} C_1
k^{3/2}-k\left[2\beta\kappa+r^2(\frac{\mu^2}{2}-\beta \Lambda)
\right]+\frac{1}{8} \left(\mu^2 r^2+2\beta\kappa \right)^2\left[[(r a)']^2 - (1-4\beta) (a^2 r)'\right]=0
\end{equation}
These two master equations, when solved with respect to $a(r)$ and $k(r)$, give a full solution to the field equations (\ref{einstein}) and (\ref{proca}) for the symmetry at hand. This is the task we will undertake in the rest of the paper. 
Already in this form we see the relation to the scalar-tensor system \cite{Babichev:2013cya} where one has $k(r)$ solving the algebraic equation (\ref{master2}) while taking $a(r)=q$. All scalar-tensor solutions are however not admitted since we still have to satisfy the modified Proca equation (\ref{master1}). This is achieved for  $\mu=\Lambda =0$ yielding the stealth Schwarzschild solution with a constant electric field and arbitrary $\beta$. An interesting twist of this solution happens for $\beta=1/4$ and the same spacetime (Schwarzschild) metric. Here, because the right hand side of (\ref{master1}) drops out, allows us to have a non-trivial electric field $a(r)=q+Q/r$ which is the interesting asymptotically flat solution found in \cite{Chagoya:2016aar}. If we allow for $\mu\neq 0$ we can get get counterpart solutions to the self tuning de Sitter black holes of \cite{Babichev:2013cya} found in \cite{Minamitsuji:2016ydr}. Again the value $\beta=1/4$ allows to have a non trivial electric field. In fact the system is completely integrable for $4\beta=1$ and we will discuss this in the next section. Finally when the Proca and Maxwell field are taken as two separate fields there will be an analogy to the system studied in \cite{Babichev:2015rva}.

Before finding explicit solutions to the above system, we will analyze the metric constraint (\ref{relfh}) in the asymptotic region $r\to\infty$. Two cases have to be distinguished,
whether the Proca mass $\mu$ vanishes or not. For $\mu\not=0$, assuming that $h$ behaves asymptotically as
some power of $r$, the constraint (\ref{relfh})  forces $f$ to have a standard asymptotic (A)dS behavior,  $f\sim r^2$. This opens the possibility of solutions that are asymptotically (A)dS or even Lifshitz. In the (A)dS case, for
$f\sim -\frac{\Lambda_{eff}}{3}r^2$ and $h\sim -\frac{\Lambda_{eff}}{3}r^2$, the metric constraint will imply that the effective cosmological constant
is fixed in term of the Proca mass as
\begin{eqnarray}
\Lambda_{eff}=-\frac{\mu^2}{2\beta}.
\end{eqnarray}
On the other hand, for $\mu=0$, if $h$ is asymptotically some power of $r$ then the metric function $f$ must go to a constant at infinity. These behaviors are in agreement with the asymptotically  flat solutions or conical geometries (for $\Lambda \neq 0$) found in \cite{Chagoya:2016aar}.

In the next two sections, we will report on two generic classes of solutions. The first one is obtained for a fixed coupling constant $1=4\beta$, and in this case, we will be able to derive the general static solution with a maximal homogeneous 2 dimensional space. These solutions can be either black holes or solitons and both are asymptotically AdS. On the latter section we will derive solutions for the case $C_1=0$. Since, as we will argue, $C_1\neq 0$ for  $1=4\beta$, this latter class of solutions will be complimentary to the former class. The class will be general for certain cases like $\mu=0$. Unlike what has been reported in the literature other asymptotically flat solutions exist in this class for other values of the coupling constant $\beta$. We will also see that Lifshitz black hole solutions can be 
obtained in this class.

\section{The general solution for  $1=4\beta$}\label{sec:special}

This case has been studied for $\mu=0$ by Chagoya et.al. \cite{Chagoya:2016aar}. After finding the asymptotically flat solution (for $\mu=\Lambda=0$) that we mentioned above, the authors argue that it is the only value of $\beta$ where we can have   such asymptotics \cite{Chagoya:2016aar}. However, as we will see in the next section, a careful analysis shows that generic cases of $\beta$ can also have asymptotically flat solutions. Our results and those of \cite{Chagoya:2016aar} are important  as (in the absence of the higher order term) a Proca vector generically spoils usual, flat, dS or adS asymptotics. In this section we concentrate on this particular value of the $\beta$ coupling where we can find the general solution and will have thus a complete picture of the static solutions in this case.
Indeed, (\ref{master1}) gives
\begin{eqnarray}
[( r\; a)']^2=\frac{Q_2^2 k(r)}{\left(\mu^2 r^2+\frac{\kappa}{2}
\right)^2} , \label{solmaster1}
\end{eqnarray}
and substituting the electric potential $a$ (\ref{master2}) gives,
\begin{eqnarray}
k(r)=\frac{1}{64 C_1^2}\,
\Big[(2\Lambda-4\mu^2)r^2+(Q_2^2-4\kappa)\Big]^2 \label{solmaster2}
\end{eqnarray}
Note that when $C_1=0$ then $k$ is undetermined and we are left with a degenerate system of one equation with two variables $a(r)$ and $k(r)$ with $\Lambda=2\mu^2$ and $Q_2^2=4\kappa$.  Taking a particular $a(r)$ will give some $k(r)$ \cite{Minamitsuji:2016ydr} but these solutions are pathological, as our analysis shows, for the system is degenerate and undetermined for this case.

Therefore from now on we stick to $C_1\neq 0$ and we see that $Q_2$ the Proca charge modifies the effective horizon curvature and may give a solid deficit angle just like for a global monopole \cite{Barriola:1989hx}. This is something we will have to keep in mind.
The $k$ function which determines the spacetime solution is now of identical form with the static $q=0$ solutions as classified in \cite{Babichev:2016rlq} where now the curvature term is replaced by $Q_2^2-4\kappa$. We will now look at this class of solutions in detail for different parameters.

Substituting the solution (\ref{solmaster1}) into (\ref{solmaster2}), one obtains,
\begin{eqnarray}
a(r)=\frac{Q}{r} - \frac{Q_2}{8\,C_1\,r}\int
\frac{\left(2\Lambda-4\mu^2\right)r^2+\left(Q_2^2-4\kappa\right)}{\mu^2
r^2+\frac{\kappa}{2}}dr
\label{inta}
\end{eqnarray}
Similarly, using (\ref{susb}) we find that,
\begin{eqnarray}
\label{inth}
h(r)=-\frac{2M}{r}+\frac{1}{(8C_1)^2r}\int \frac{\Big[(2\Lambda-4\mu^2)r^2+(Q_2^2-4\kappa)\Big]^2}{\mu^2 r^2+\frac{\kappa}{2}} dr
\end{eqnarray}
Both of these integrals can be easily found depending on the value of
$\kappa=0,\pm 1$.

\subsection{The spherically symmetric adS black holes and solitons}
\label{ssec:solitons}
For spherical symmetry, the electric potential $a(r)$ is given by
\begin{eqnarray}
a(r)=\frac{Q}{r} +\frac{Q_2}{2 C_1\mu^3
}\Big[-\frac{\sqrt{2}}{4}\left((Q_2^2-2)\mu^2-\Lambda\right)\frac{\arctan(\mu\sqrt{2}
r)}{r}+\mu\left(\mu^2-\frac{\Lambda}{2}\right)\Big],
\label{ak}
\end{eqnarray}
and we have two integration constants, $Q$ alike Coulomb charge and $Q_2$.
The metric function $h$ takes the following form,
\begin{equation}
\begin{split}
h(r) & =\frac{(2Q_2^2\mu^2-6\mu^2-\Lambda)(\Lambda-2\mu^2)}{2(4 C_1 \mu^2)^2}-\frac{2M}{r}\\
 &-\Lambda_{eff}\Big[\frac{r^2}{3}+\frac{(Q_2^2\mu^2-2\mu^2-\Lambda)^2}{2\sqrt{2}\mu^3r(\Lambda-2\mu^2)^2}\arctan{(\sqrt{2}r \mu)}\Big],
\end{split}
\label{hk}
\end{equation}
where we set,
\begin{eqnarray}
\Lambda_{eff}=-\Big[ \frac{\Lambda-2\mu^2}{4 C_1 \mu}\Big]^2
\end{eqnarray}
for the effective cosmological constant and for the effective horizon curvature term. The solution depends on four integration constants $Q, Q_2, C_1$ and $M$. 
The latter charge $M$, is part of the overall mass, since the arctan term in (\ref{hk}) contributes similarly at asymptotic infinity. 
The constant $C_1$ is not physical, as it corresponds to the reparametrisation of time, i.e. the gauge choice. Later we will fix it such that at infinity we recover the 
standard form of adS metric.
$Q$ is the Coulomb charge which is a stealth parameter for the spacetime solution. In EM theory this electric charge would give rise to the RN black hole solution. This stealth feature is a particular feature associated to $\beta=1/4$ as part of the $a$ dependence in (\ref{master2}) drops off from the field equations. The $Q_2$ charge on the other hand is related to the breaking of gauge symmetry due to the Proca mass term.
Secondly, we remark that the effective cosmological constant is fixed and always negative for $\mu^2>0$. Finally, we note the presence of the latter $\arctan$ over $r$ term in both $h$ and $a$. This key term will contribute a finite number at $r=0$ and as a result will influence the regularity of the solution. Indeed we can easily see that if $M=0$,
\begin{equation}
\label{hr0}
h(0)=2 \left(\frac{4-Q_2^2}{8 C_1 }\right)^2,
\end{equation}
which is always positive or zero.  
Last but not least we have,
\begin{eqnarray}
f(r)=2 h(r)\left(\frac{4 C_1 \mu^2}{\Lambda-2\mu^2}\right)^2 \left(\frac{r^2+\frac{1}{2\mu^2}}{r^2+\frac{Q_2^2-4}{2\left(\Lambda -2\mu^2\right)}}\right)^2.
\label{fk}
\end{eqnarray}
and we can see that the effective curvature is always equal to unity, $f(0)=1$ for $M=0$, making curvature regular at $r=0$. Therefore these solutions are always locally adS and there is no solid deficit angle.

In order to have the standard form of asymptotically adS solutions we have to fix $C_1$ so that,
\begin{equation}
\label{C1fix}
2\left(\frac{4 C_1 \mu^2}{\Lambda-2\mu^2}\right)^2=1.
\end{equation}
This ensures an identical behavior for $f$ and $h$ for large $r$ and it is equivalent to fixing the gauge.
In this case the effective cosmological constant is given by the Proca mass parameter since,
\begin{eqnarray}
\Lambda_{eff}=-2\mu^2.
\end{eqnarray}
The resulting solution is always an asymptotically adS black hole. It has very similar properties to the spherical or planar adS static black holes depending on the value of $Q_2$. This is because the latter $\frac{\arctan}{r}$ term in (\ref{hk}) is everywhere bounded, finite at $r=0$ and decays at infinity as a mass term with a $1/r$ falloff. Again, we emphasize that the usual Coulomb charge $Q$ does not influence the spacetime metric.

Let us look into more detail the solution for $M=0$ and $Q_2\neq 2$. The solution has no solid angle deficit as we have $f(0)=1$. A solid angle deficit (or excess) would have meant that spacetime is singular for $r=0$ (even if $h$ is regular there). Here, we have the nice result that the metric is completely regular and hence for $M=0$ we have a regular soliton solution for arbitrary Proca mass $\mu$ which has asymptotic adS geometry.  We also see here that the addition of the curvature-vector interaction term, $G^{\mu\nu}A_\mu A_\nu$ smooths the effects of the Proca mass term giving a regular solution with adS asymptotics. When we switch on the mass we have a black hole (with adS asymptotics). This is radically different from an electrically charged RN black hole where the $M=0$ spacetime is actually singular. It would seem that in Proca theory and for $\beta=1/4$, when the Proca mass is corrected by  curvature interaction the situation is regularized.   
The full spacetime solution, with mass $M$ included and fixed $C_1$ as in (\ref{C1fix}), reads,
\begin{eqnarray}
h(r)=\frac{2\mu^2}{3} r^2+\frac{2Q_2^2\mu^2-6\mu^2-\Lambda}{\Lambda-2\mu^2}-\frac{2M}{r}
+\frac{(Q_2^2\mu^2-2\mu^2-\Lambda)^2}{\sqrt{2}\mu (\Lambda-2\mu^2)^2}\frac{\arctan{(\sqrt{2}r \mu)}}{r},\nonumber\\
\label{hks}
\end{eqnarray}
\begin{eqnarray}
f(r)=h(r)\left(\frac{r^2+\frac{1}{2\mu^2}}{r^2+\frac{Q_2^2-4}{2\left(\Lambda -2\mu^2 \right)}}\right)^2.
\label{fks}
\end{eqnarray}
The Proca charge $Q_2$ is associated to the breaking of $U(1)$ gauge invariance and it can take arbitrary values. 
There are however particular values of $Q_2$, for which the solution is special. 
In particular the last term in~(\ref{hks}) drops out, if 
\begin{equation}
Q_2^2 \mu^2-2\mu^2-\Lambda=0.
\end{equation}
With this choice we get a stealth Schwarzschild-adS  solution \cite{Minamitsuji:2016ydr},
\begin{equation}
h(r)=\frac{2\mu^2}{3} r^2+ 1-\frac{2M}{r}.
\end{equation}
On the other hand for 
$$2Q_2^2\mu^2-6\mu^2-\Lambda=0$$
the effective curvature is zero although we have a spherical horizon. 
For other values of $Q_2$ we have a non stealth solution which for $M=0$ becomes a soliton\footnote{Note that $Q_2=\pm2$ must be discarded in the case $M=0$, since the solution is singular at $r=0$, see Eq.~(\ref{hr0})}. Therefore there are three distinct sub-classes of solutions with adS asymptotic within this class.

Note that although in the case of the soliton $M=0$, the mass of the soliton is not zero. Indeed, one can deduce the soliton mass from the asymptotic 
behavior at large $r$ in~(\ref{hks}).  For $r\to \infty$ the last term has the form $\sim 1/r$, therefore one can formally define the effective mass of the soliton as 
\begin{equation}
	\label{Meffsol}
	M_{eff} = - \frac{\pi(Q_2^2\mu^2-2\mu^2-\Lambda)^2}{2\sqrt{2}\mu (\Lambda-2\mu^2)^2}.
\end{equation}

There are a number of special values for the coupling constants.
Choosing $\Lambda=2\mu^2$ will kill the effective cosmological constant $\Lambda_{eff}=0$. We get,
\begin{equation}
h(r)=-\frac{2M}{r}+\sqrt{2\mu^2}\frac{(Q_2^2-4)^2}{(8 C_1 \mu)^2}\frac{\arctan{(r \sqrt{2\mu^2})}}{r}
\end{equation}
The solution has generically an event horizon. However it also has unusual asymptotics as $h\rightarrow 0$ while $f \sim r^4$ for $r\rightarrow +\infty$. Another particular limit is to take $\mu=0$ and this solution was found in \cite{Chagoya:2016aar}. We note that in presence of a $\Lambda-$term the solution behaves asymptotically
as a conical geometry \cite{Cvetic:2014nta}
$$
ds^2\sim -r^4dt^2+5dr^2+r^2d\Omega^2.
$$
for large $r$. For $M=0$ however we see that $f(0)=1$ and therefore the solution has regular curvature at $r=0$. This agrees with the result of \cite{Chagoya:2016aar}. Asymptotically however space will have a solid deficit angle removed from the sphere similar to the global monopole solution \cite{Barriola:1989hx}.

If we additionally set $\Lambda=0$, the solution becomes,
$$
h(r)=1-\frac{2M}{r}\frac{32 C_1^2}{(4-Q_2)^2},\qquad f(r)=\frac{32 C_1^2}{(4-Q_2)^2} h(r),
$$
In order to make it asymptotically flat, the constant $C_1$ is fixed in terms of $Q_2$ as
$$
C_1^2=\frac{1}{32}(4-Q_2)^2
$$
and this explains why the asymptotically flat solution reported in \cite{Chagoya:2016aar} has only three integration constants.

It is easy to see that choosing hyperbolic geometries, $\kappa=-1$, the arctangent term in (\ref{hk}) will be replaced by a hyperbolic arctangent which will explode exponentially at finite $r$. These solutions can be trivially obtained but will have singular asymptotics, and hence we do not discuss them further. To get an arctangent term and de Sitter asymptotics for $\kappa=-1$ one could consider an imaginary Proca mass term, $\mu^2=-m^2$. Although such a term would be discarded due to instability in usual Proca theory, here, the presence of higher order terms does not guarantee this intuition. However, a quick analysis in this case shows that the solution has always negative effective curvature and as a result is always singular for de Sitter asymptotics.

\subsection{Planar horizon black holes $\kappa=0$}
Let us suppose now that the horizon's geometry is locally flat, $\kappa=0$,
$$
ds^2=-h(r)dt^2+\frac{dr^2}{f(r)}+r^2(dx^2+dy^2),\qquad
$$
Integrating (\ref{inta}) it is straightforward to obtain the electric potential,
\begin{eqnarray}
 a(r)=\frac{Q}{r}-\frac{Q_2}{8C_1\mu^2
r^2}\Big[\left(2\Lambda-4\mu^2\right)r^2-Q_2^2\Big].
\end{eqnarray}
whereas from (\ref{inth}) the metric functions take the form,
\begin{eqnarray}
h(r)&=&\left(\frac{2\mu^2-\Lambda}{4 C_1 \mu}\right)^2\frac{r^2}{3}+\frac{(\Lambda-2\mu^2)Q_2^2}{(4 C_1 \mu)^2}
-\frac{2M}{r}-\frac{Q_2^4}{(8 C_1\mu)^2 r^2}\\
f(r)&=& \frac{128\mu^4 C_1^2 r^4}{\left(4\mu^2 r^2-2\Lambda r^2-Q_2^2\right)^2}h(r).
\end{eqnarray}
Although we have a planar geometry for the horizon surface, the black hole potential is similar to that of an adS RN geometry, however, with imaginary charge. Additionally, the effective curvature term is of an undetermined sign fixed by the Lagrangian parameters,  while the effective cosmological is always negative. The imaginary charge term means that even at the absence of mass $M$ the central singularity will be dressed by an event horizon since $h'$ is positive. This is contrary to the usual RN solution which is singular for small black holes. The asymptotics are locally adS.

In order to have adS asymptotics as before we must impose,
\begin{equation}
\left[\frac{4C_1 \mu}{\Lambda-2 \mu^2}\right]^2=\frac{1}{2\mu^2}
\end{equation}
effectively fixing $C_1$ and we get the solution,
\begin{equation}
h(r)=\frac{2Q_2^2 \mu^2}{\Lambda-2\mu^2}+r^2 \frac{2\mu^2}{3}-\frac{2M}{r}-\frac{(Q_2^2 \mu)^2}{2 r^2(\Lambda-2\mu^2)^2}
\end{equation}
\begin{equation}
f(r)=\frac{h(r)}{1+\left(\frac{Q_2^2}{(2\Lambda-4 \mu^2)r^2}\right)}
\end{equation}
Again we see that even if $M=0$ we have a black hole horizon dressing the singularity at $r=0$. This is due to the Proca charge which now is of the form of an imaginary RN charge. We can have an effective positive or negative curvature term depending on the sign of $\Lambda- 2 \mu^2$ but it does not change the properties of the solution. This is because it is always the imaginary charge that is dominant at smaller $r$.

\section{Solutions for $C_1=0$ and arbitrary $\beta$.}\label{sec:beta}

For the special coupling $\beta=\frac{1}{4}$, we are able to obtain the general spherical, hyperbolic or planar, static
solution. 
In order to obtain the general solution for arbitrary $\beta$ one would have to resort to some numerical integration. Indeed one would solve for $k$ from (\ref{master2})
and then numerically solve for $a$ using (\ref{master1}). We will not undertake this task here. We saw also that the constant $C_1$ could not be set to zero for $\beta=\frac{1}{4}$. Therefore it is instructive and easier to study now the particular case of $C_1=0$ while keeping the coupling $\beta$ arbitrary. We will see the asymptotic nature of the solutions does not generically change and we again obtain asymptotically flat of adS solutions. In order to achieve the solution of the system for $C_1=0$,  it is convenient to bring down the order of the
equation (\ref{master1}) by introducing the variables $X$ and $y$ as follows,
\begin{equation}
\label{X0}
X= \frac{k}{(\mu^2 r^2+2\beta\kappa)^2},
\end{equation}
The metric function $h(r)$ (\ref{susb}) is now given by,
\begin{eqnarray}
h(r)=-\frac{2M}{r}+\frac{1}{r}\int X(r)(\mu^2 r^2+2\beta\kappa) dr
\label{susb1}
\end{eqnarray}
On the other hand we set
\begin{equation}
\label{y}
y=\frac{a}{r a'}.
\end{equation}
Indeed, in this case, after straightforward calculations, the master equations we have to solve, (\ref{master1}) and (\ref{master2}), reduce to,
\begin{equation}
\label{master11}
 \frac{X'}{2X}=\frac{1+y-ry'}{ry(1+4\beta y)},
\end{equation}
\begin{equation}
\label{master21} C_1 \left(\mu^2 r^2+2\beta\kappa \right) X^{3/2}- X\left[2\beta\kappa+r^2(\frac{\mu^2}{2}-\beta \Lambda)
\right]
+\frac{a^2}{8y^2} \left(1+8\beta y +4 \beta y^2 \right)=0.
\end{equation}
Eqs.~(\ref{y}), (\ref{master11}) and (\ref{master21}) form a closed system of equations on three functions: $y$, $X$ and $a$. Notice that if $a$ is any power of $r$ then $y$ is just a constant given by the power in question. With this observation one can verify that any constant $y$ cannot yield a solution to the system except in particular cases like $\kappa=0$ (see the Lifshitz section). 
Finally, if $a$ is constant then $1/y$ is exactly zero, and the resulting solution is nothing but the stealth configuration on the Schwarzschild AdS spacetime \cite{Babichev:2013cya}. 

For the moment we have not achieved much from the change of variables, as the field $a$ does not completely cancel out in the above system of equations; unless $C_1=0$. Indeed, in this latter case,
 using (\ref{master21}) we find $X$ and replace it in (\ref{master11}) to get,
\begin{eqnarray}
\frac{dr}{r}\left[(1-4\beta)+ \frac{r^2(\mu^2-2\beta\Lambda)}{4\beta\kappa+r^2(\mu^2-2\beta\Lambda)}(1+4\beta y)\right]
-\frac{4\beta (1-4\beta)ydy}{1+8\beta y+4\beta y^2}=0.
\label{eqsep}
\end{eqnarray}
In addition to the vanishing $\beta$ case\footnote{Indeed, from Eqs. (\ref{master11}-\ref{master21}), it is easy to see that pure Proca theory $\beta=0$ with $C_1=0$ gives an unphysical metric. Nevertheless, in the case where $C_1\not=0$, numerical solutions have been reported in \cite{Liu:2014tra}.}, there are three generic cases given by $\mu=\Lambda=0$, $\mu^2=2\beta \Lambda$ and finally $\kappa=0$ for which the above equation is separable. In these cases, for $C_1=0$, the full system reduces to a single ODE (\ref{eqsep}) and the full solution is known for arbitrary $\beta$. In what follows we will discuss each of the cases in detail, first the former two and then the latter $\kappa=0$. We can already anticipate the form of $f$ by noting,
$$
f(r)=\frac{h(r)}{2\beta X}
$$
while,
\begin{equation}
\label{X}
X(r)=\frac{\beta a^2}{y^2}\frac{y^2+2y + \frac{1}{4\beta}}{4\beta \kappa+r^2 (\mu^2-2\beta \Lambda)}
\end{equation}

\subsection{The case $\kappa=1$, for $\mu=\Lambda=0$ or $\mu^2=2\beta \Lambda$}
The assumption $\mu^2=2\beta \Lambda$ greatly simplifies Eq.~(\ref{eqsep}), yielding,
\begin{equation}\label{talca}
	\left(\frac{1}{4\beta}+2 y +y^2\right)\frac{dr}r =  y dy 
\end{equation}
First of all we see that "power of $r$" solutions are excluded. Indeed then $y$ is a constant and is fixed by the requirement $\frac{1}{4\beta}+2 y +y^2=0$. However then we see that $X=0$ from (\ref{X}) which is not allowed. Otherwise we have a separable ODE which can be explicitly solved by coordinate transforming from $r$ to $y$ coordinates.  
The effective expression in terms of $y$ depends on the solutions of the quadratic equation $1+8\beta y +4\beta y^2=0$. 
Namely for $4\beta>1$ or $\beta<0$ we will have two real roots while for $0<\beta<1/4$ there are no real roots. Each root corresponds to an endpoint of spacetime, either asymptotic infinity in $r$ or a singularity. Therefore, each interval in $y$ will map to a different solution in $r$ coordinates.  Since we are seeking black hole solutions we will seek regions in $y$ coordinates where $r$ can asymptote infinity. 
Also note that the case $\beta=1/4$ is not covered by Eq.~(\ref{talca}), since we divided by the factor $(1-4\beta)$ to get (\ref{talca}),
and the case $\beta=0$ does not give physically consistent solutions, as we mentioned before. Let us now solve step by step and determine the region we want to study. 

Using~(\ref{talca}) and (\ref{y}) we can also find a differential equation on $a$ as a function of $y$ variable,
\begin{equation}\label{ay}
	\frac{d a}a = \frac{4\beta\; dy}{1+8\beta y + 4 \beta y^2}.
\end{equation}
We can express the physical metric and the vector field in terms of the new coordinate $y$. 
In particular, from Eq.~(\ref{susb}) we obtain,
\begin{equation}
\label{hofy}
h(r)=-\frac{2M}{r}+\frac{1}{r}\int \frac{r a^2}{4\kappa}\frac{dy}{y} (\mu^2r^2+2\beta\kappa),
\end{equation}
where $r$ and $a$ are now understood as explicit functions of $y$ from (\ref{ay}) and (\ref{talca}).

Let us assume now that $4\beta>1$ and set , 
\begin{equation}
\gamma=\sqrt{\frac{4\beta-1}{4\beta}}.
\end{equation}
so that $\frac{1}{4\beta}+2 y +y^2=(y+1+\gamma)(y+1-\gamma)$. Note that for our choice of $\beta>1/4$, we have that $0<\gamma<1$ and the roots are both negative.
Then a straightforward integration of (\ref{talca}) gives,
\begin{equation}\label{rofy}
\frac{r}{r_0} = \frac{(y+1+\gamma)^\frac{1+\gamma}{2\gamma}}{(y+1-\gamma)^\frac{1-\gamma}{2\gamma}},
\end{equation}
and we see that when $r\rightarrow +\infty$ then $y\rightarrow +\infty$. 
While integrating~(\ref{ay}) we find,
\begin{equation}\label{aofy}
\frac{a}{a_0} = \left(\frac{y+1-\gamma}{y+1+\gamma}\right)^\frac{1}{2\gamma},
\end{equation}
where $r_0$ and $a_0$ are constants of integration. It is now straightforward to see that
\begin{equation}
f(y)=\frac{h(y) y^2}{2 \beta a_0^2}\frac{(y+1+\gamma)^{\frac{1-\gamma}{\gamma}}}{(y+1-\gamma)^{\frac{1+\gamma}{\gamma}}},\qquad \frac{dr^2}{f(y)}=2\beta a_0^2 r_0^2 \frac{dy^2}{h(y)}
\end{equation}
which shows that the metric acquires a homogeneous form in $y$-coordinates.  
Substituting the expressions (\ref{rofy}) and (\ref{aofy}) into (\ref{hofy}), we obtain for the metric function $h$,
\begin{equation}
\label{hofy2}
h(y)=\frac{1}{r(y)}\left(-2M
+\frac{\beta r_0 a_0^2}{2 } I_1(y)
+\frac{r_0^3 a_0^2 \mu^2}{4\kappa} I_2(y)\right),
\end{equation}
where we introduced the notation,
\begin{equation}
	I_1 = \int \frac{dy}{y} \frac{(y+1-\gamma)^\frac{1+\gamma}{2\gamma}}{(y+1+\gamma)^\frac{1-\gamma}{2\gamma}},\;\;\;
	I_2 = \int \frac{dy}{y} (y+1-\gamma)^\frac{3\gamma-1}{2\gamma} (y+1+\gamma)^\frac{3\gamma+1}{2\gamma}.
\end{equation}
Since $y=0$ is a singularity for the integrals the interval we want to focus on is $y>0$. The remaining singularities of the integrals are excluded from this region for $0<\gamma<1$. 

The integration constant $r_0$ simply rescales the radial coordinate, while $a_0$ is related to gauge choice. Indeed for large $y$ we find that $a\to a_0$, as it can be seen from~(\ref{aofy}).
From~(\ref{rofy}) we also notice that the coordinate $y$, covers not all $r$, but
$r\geq r_{min}\equiv \frac{3\sqrt{3}}2r_0$. 
As we will see below, this is not a problem, since $r_{min}$ corresponds to a singularity hidden by an event horizon. The solutions are always asymptotically flat for $\mu=0$ while they are adS for $\mu\neq 0$. In order to get an explicit solution we can set $\frac{1+\gamma}{2\gamma}=n$ and take $n$ to be a positive integer greater than $1$. Simple integration then shows that the above integrals are finite power series of $n$. For example $I_1$ takes the generic form,
\begin{equation}
I_1=y+a_0 \ln\frac{y}{(2n-1)y+2n} + \sum_{k=0}^{k=n-2}  \frac{(-1)^k a_k}{\left((2n-1)y+2n\right)^k}
\end{equation}
where $a_i$ are some numerical coefficients. Note then that for $y$ big enough the solution asymptotes a Schwarzschild solution and is therefore asymptotically flat with the correct Newtonian falloff.
Taking $n=2$ and $\mu=0$ for example gives us,
\begin{equation}
\label{h0}
h_0(y)=\frac{1+\frac{2}{3y}}{(1+\frac{4}{3y})^2}\left(1+\frac{1}{3y}\ln \frac{y}{3y+4}-\frac{2M}{y}\right)
\end{equation}
giving an asymptotically flat black hole (even with $M=0$). Indeed fixing $2\beta r_0^2 a_0^2=1$ gives $f=h$. 

Switching on Proca mass $\mu\neq 0$ we get instead adS asymptotics in the same fashion as in the previous section. Indeed we have
\begin{equation}
h_\mu(y)=h_0(y)+\frac{1+\frac{2}{3y}}{(1+\frac{4}{3y})^2}  \left(1+\frac{6}{y}+\frac{16}{y^2}+\frac{64}{9} \ln y  \right)\frac{\mu^2}{3} y^2 
\end{equation} 
and Proca mass plays the role of an effective cosmological constant. Higher powers of $n$ will yield higher negative powers in the expression for $h$ but the solution is always asymptotically flat or adS for $\mu=0$ and non zero respectively. For example, expression (\ref{h0}) picks up an extra $1/y^2$ for $n=3$ and then an additional $1/y^3$ term for $n=4$ etc...
It is interesting to note that as $n$ goes to infinity then $\beta$ attains the special value $1/4$ while as $n$ tends to $1$ we get that $\beta$ goes to infinity.

\begin{figure}[tbp]
\centering 
\includegraphics[width=.45\textwidth]{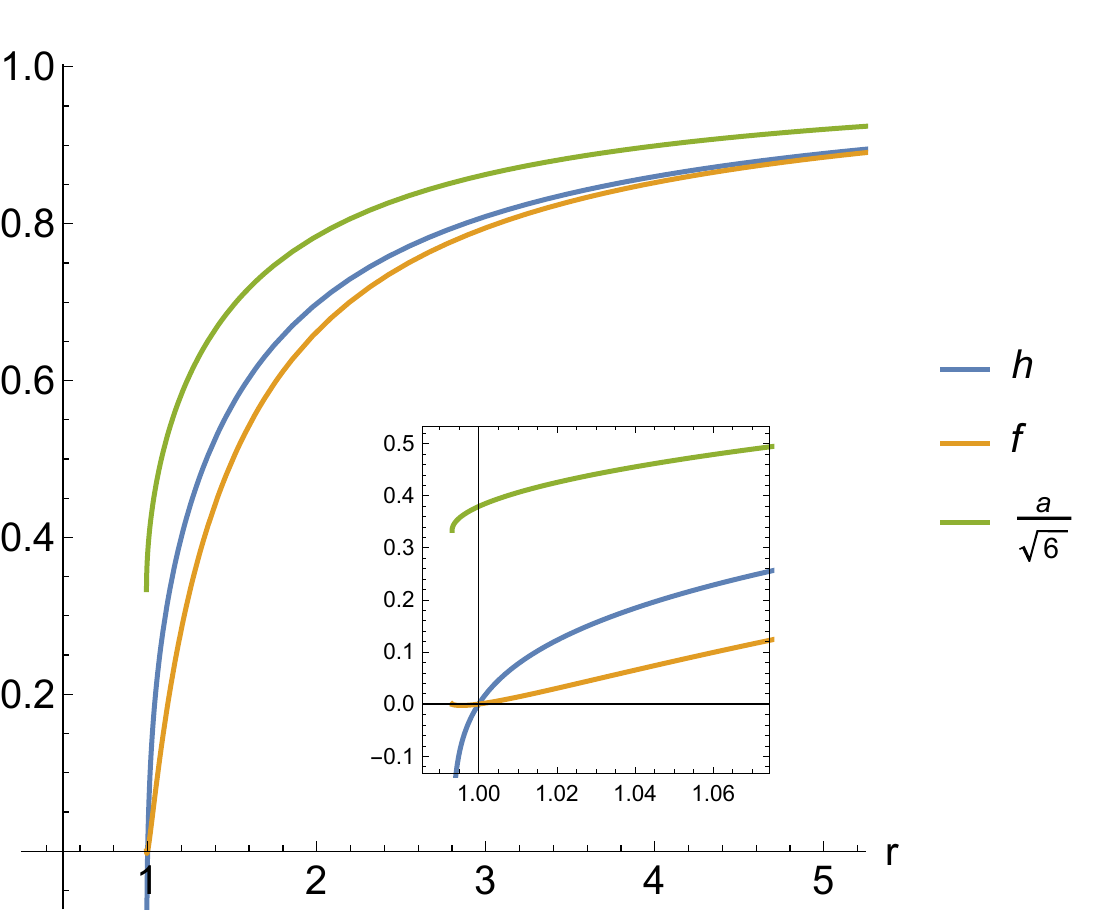}
\hfill
\includegraphics[width=.45\textwidth,origin=c]{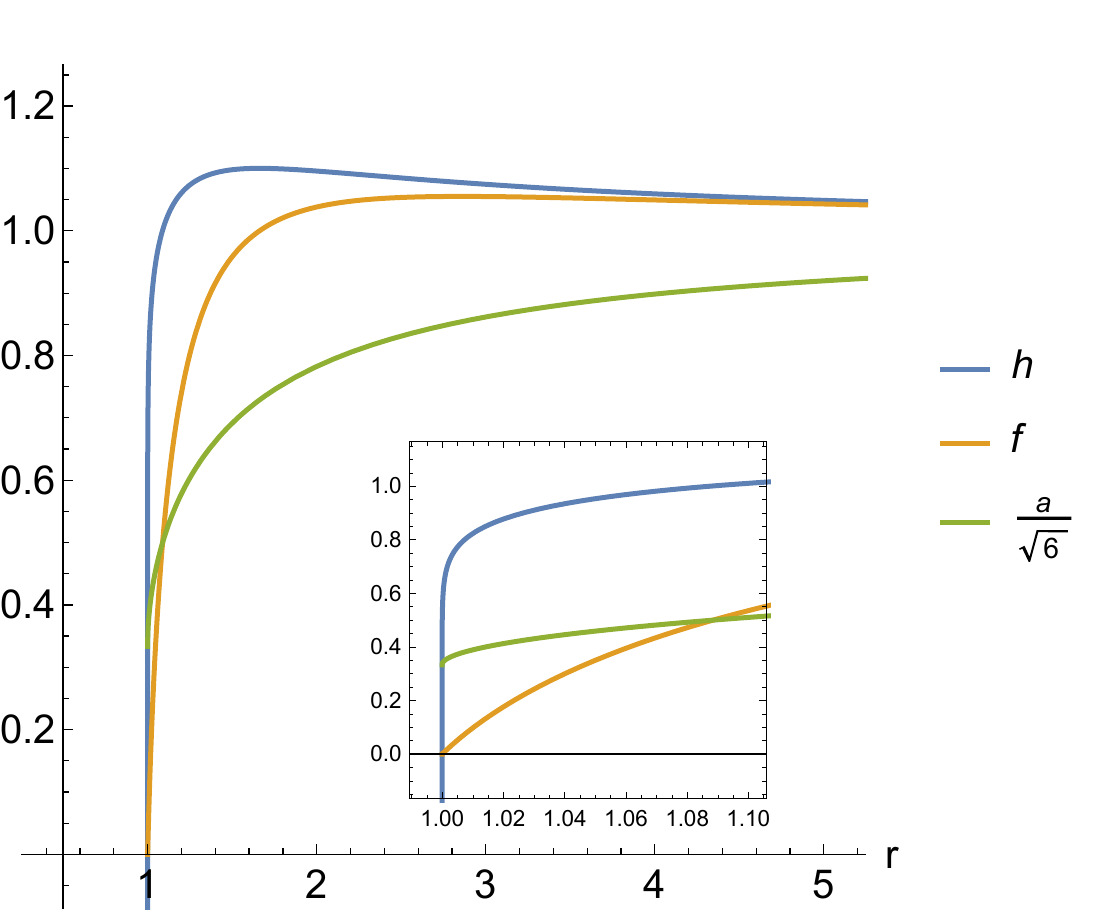}
\caption{\label{fig:sol1} Black hole solutions for $\mu=0$ and a) $M=0$ (the left panel); b) for negative $M$, such that the effective mass at infinity is negative (the right panel).}
\end{figure}

It is possible to work out other explicit solutions for $n$ non integer, for example, for the value of $\beta=1/3$.
The details of the integral calculation are presented in Appendix~\ref{app int}, and the result is depicted in~Fig.~\ref{fig:sol1}.
Fixing the gauge such that $h(\infty)=1$, we find the asymptotic behaviour as follows,
\begin{equation}
	h \simeq f\simeq 1 - \frac{M_{eff}}{r}, \quad r\to \infty,
\end{equation}
where
\begin{equation}
	M_{eff} = M+r_0\left(1-\frac{\log(2-\sqrt{3})}{2\sqrt{3}}\right).
\end{equation}
Note that even for $M=0$, the event horizon exists, see the left panel of~Fig.~\ref{fig:sol1}. 
It is interesting that choosing negative value for the bare mass $M$, there is a black hole solution with negative asymptotic effective mass, $M_{eff}<0$,
see the right panel of~Fig.~\ref{fig:sol1}.
The horizon does exist in this case, but the far away observer would measure repulsive gravitational force. 
We also would like to stress that the curvature singularity taking place at $y=0$ is always hidden behind the event horizon.

\begin{figure}[t]
\centering 
\includegraphics[width=.5\textwidth]{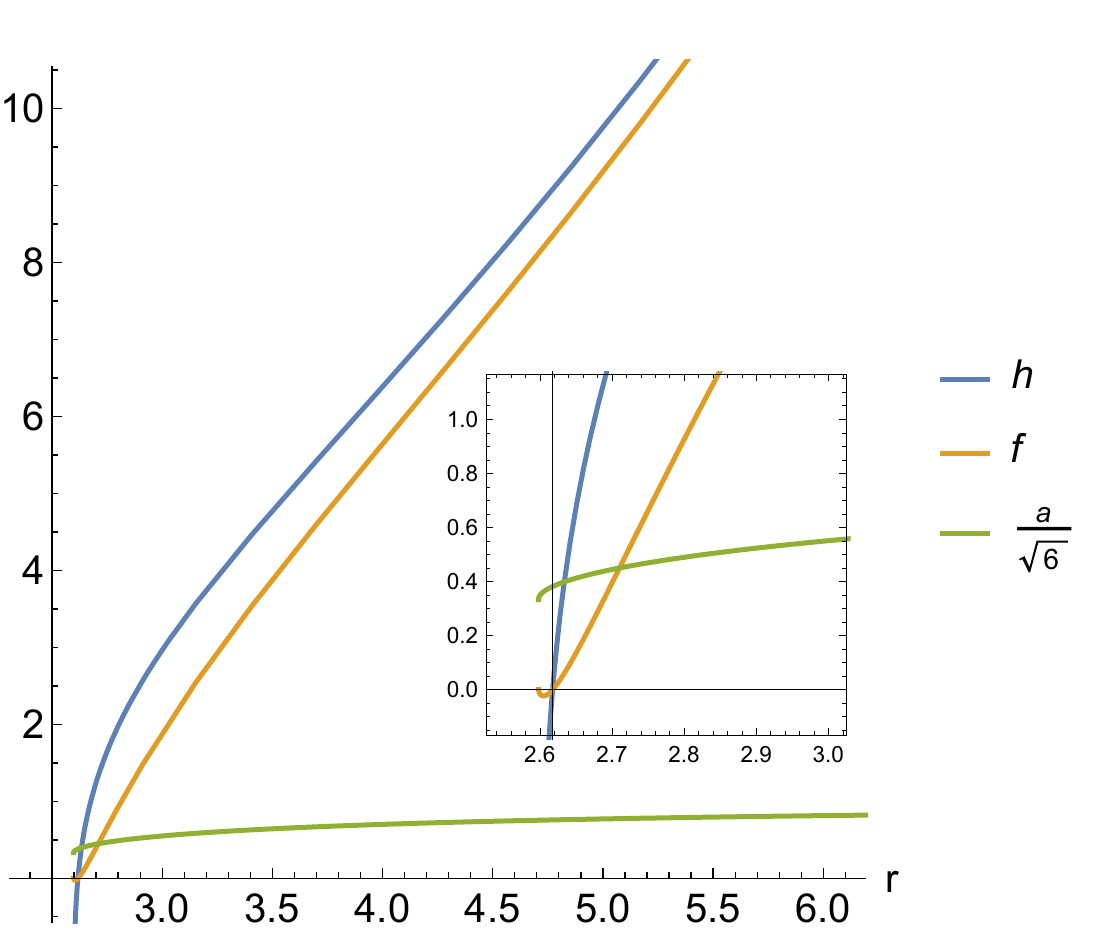}
\caption{\label{fig:sol2} Black hole solution for $M=0$ and non-zero $\mu$ , leading to the AdS asymptotic behaviour.}
\end{figure}

A black hole solution for non-zero $\mu$ is shown in Fig.~\ref{fig:sol2}. As in the case $\mu=0$ the singularity is covered with the event horizon.
The effect of nonzero $\mu$ is the AdS asymptotic behaviour at $r\to \infty$.

\subsection{Lifshitz black holes: Topological $\kappa=0$ case with $C_1=0$ }

We have already found analytical solutions with a planar base manifold $\kappa=0$ corresponding to asymptotically AdS black holes in the
$4\beta=1$ case. Here, for $C_1=0$, the equation (\ref{eqsep}) becomes separable yielding
\begin{equation}
\label{litsi}
\left[1-2\beta+2\beta y \right]\frac{dr}{r}=\frac{(1-4\beta) ydy}{2\left[  \frac{1}{4\beta}+2 y +y^2 \right]}.
\end{equation}
Unlike in the previous subsection here, $1-2\beta+2\beta y=0$ yields constant $y$ solutions.  They in fact correspond to Lifshitz
spacetimes which we will turn to now.

In the last decade, there has been some interest in extending the ideas underlying the AdS/CFT correspondence to field theories with an
anisotropic scaling symmetry. In analogy with the AdS/CFT correspondence, the gravity dual metric, commonly known as Lifshitz spacetime, enjoys a scaling symmetry 
for which the spatial and temporal coordinates scale with different weight. Because of this anisotropy, Lifshitz spacetimes in
contrast with AdS, can not be sustained by pure Einstein gravity with eventually a cosmological constant, and instead require the introduction of source or higher-order
gravity theories. From the advent of Lifshitz spacetimes, it was clear that a massive Proca field coupled to Einstein gravity with a
negative cosmological constant constitutes an excellent toy model to  engineer Lifshitz spacetime \cite{Taylor:2008tg}. On the other hand, the massive character of the Proca field in spite of being
completely compatible with the Lifshitz asymptotics may be a strong obstruction to generate black holes. In fact, to our knowledge, all the Lifshitz black
hole solutions found in the literature for the Einstein-Proca model require the introduction of an extra source materialized by a $U(1)$ gauge field \cite{Pang:2009pd, Alvarez:2014pra}. In
the present case, we will see that the nonminimal coupling which plays the role of a mass term, $G_{\mu\nu}A^{\mu}A^{\nu}$, permits the emergence of Lifshitz black holes without
the need of additional fields. 

Lifshitz spacetimes take the form,
\begin{eqnarray}
\label{Lifshitz}
ds^2=-r^{2z}Fdt^2+\frac{dr^2}{r^2F}+r^2\left(dx_1^2+dx_2^2\right),
\end{eqnarray}
where $z$ denotes the dynamical exponent that is responsible for the anisotropy ($z=1$ is adS as a special case). The metric function $F$ is such that $\lim_{r\to\infty}F(r)=1$ in order
to ensure the correct Lifshitz asymptotic. 
For our case here, setting constant $y=\frac{2\beta-1}{2\beta}$  in (\ref{master11}) one finds that $X=X_0 r^{\frac{2}{2\beta-1}}$, where $X_0$ is an integration constant. 
The constant $X_0$ is fixed in such a way that $h$ has the Lifshitz spacetime, $X_0 = \frac{6\beta-1}{2\beta-1}$, so that
\begin{equation}
\label{F}
F(r)=1-\frac{2M}{r^{2z+1}},
\end{equation}
where the Lifshitz exponent is given by 
\begin{equation}
\label{z}
z=\frac{2\beta}{2\beta-1}.
\end{equation}
On the other hand the metric constraint (\ref{relfh}) with $\kappa=0$ imposes a constraint on $\mu$, 
\begin{equation}
\label{mu}
\mu^2 = 2\beta(2z +1).
\end{equation}
The full metric then takes the form~(\ref{Lifshitz}) with $z$ given by~(\ref{z}):
\begin{equation}
\label{LifshBH}
ds^2=-r^{2z}\left(1-\frac{2M}{r^{2z+1}}\right)dt^2+
\frac{dr^2}{r^2\left(1-\frac{2M}{r^{2z+1}}\right)}+r^2(dx_1^2+dx_2^2),
\end{equation}
The Proca field reads then,
\begin{eqnarray}
a(r)=\pm\frac{r^z}z\sqrt{\frac{2(\mu^2-2\beta\Lambda)}{\left(4\beta-1\right)(3\beta-1)}},
\end{eqnarray}
with $z$ and $\mu$ given by~(\ref{z}) and (\ref{mu}) correspondingly.
Note that the mass term in the Lifshitz metric (\ref{Lifshitz}) will decay only for $\beta\in ]-\infty,\frac{1}{6}[ \cup ]\frac{1}{2},\infty[$
excluding the option  $\beta=1/4$.

In the general case where $y$ is not constant we have to proceed as we did in the last section. Here we take for simplicity $\gamma^2>1/2$. We start by resolving (\ref{litsi}) while coordinate transforming $y=1/b-1+2\gamma^2$ thus obtaining in turn,
\begin{equation}\label{litsi2}
\frac{r(b)}{r_0} = \left(1+\gamma(2\gamma-1)b \right)^\frac{\gamma-1}{2(2\gamma-1)}\left(1+\gamma(2\gamma+1)b\right)^\frac{\gamma+1}{2(2\gamma+1)},
\end{equation}
Note that we have chosen the coordinate $b$ so that it has the same asymptotic behavior as $r$ for large $r$.
Similarily,
\begin{equation}\label{litsi3}
\frac{a(b)}{a_0} = \left(1+\gamma(2\gamma-1)b\right)^\frac{1}{2(2\gamma-1)}\left(1+\gamma(2\gamma+1)b\right)^\frac{-1}{2(2\gamma+1)},,
\end{equation}
In order to obtain the metric we need to coordinate transform to $b>0$ coordinates,
\begin{equation}
\label{hk0}
h(b)=-\frac{2M}{r(b)}+\frac{r_0 a_0^2 \mu^2(4\beta-1)}{4(\mu^2-2\beta \Lambda)r(b)} I_3
\end{equation}
where 
\begin{equation}
I_3=\int d b \; \frac{(1+\gamma(2\gamma-1)b)^\frac{\gamma+1}{2(2\gamma-1)}(1+\gamma(2\gamma+1)b)^\frac{\gamma-1}{2(2\gamma+1)}}{1+(2\gamma^2-1)b}
\end{equation}
The solution asymptotes the Lifshitz solution~(\ref{LifshBH}) for large $b$, provided that we fix $\mu$ as in~(\ref{mu}) and the integration constants $r_0$ and $a_0$ in the following way,
\begin{equation}
\frac{a_0}{r_0^z}=\frac1z \sqrt{\frac{2(\mu^2-2\beta\Lambda)}{\left(4\beta-1\right)(3\beta-1)}}\left(\frac{2\gamma+1}{2\gamma-1}\right)^{\frac{\gamma}{2(2\gamma^2-1)}}.
\end{equation}
In general the integration can be performed numerically. 

\begin{figure}[tbp]
\centering 
\includegraphics[width=.48\textwidth]{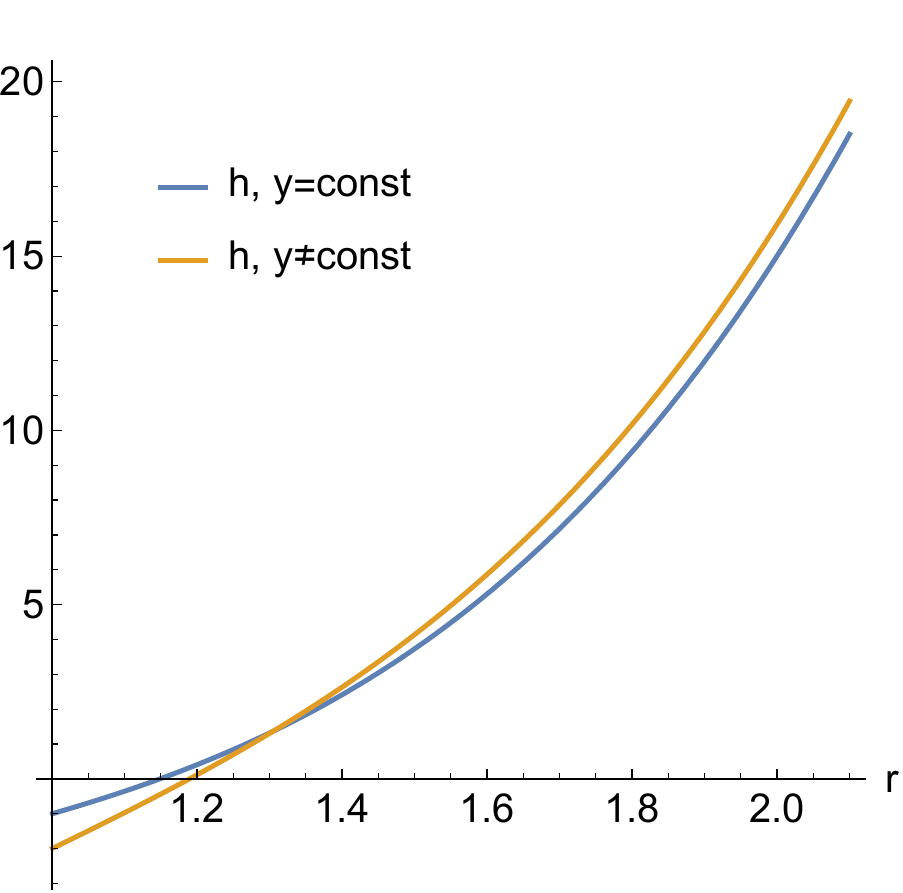}
\hfill
\includegraphics[width=.45\textwidth,origin=c]{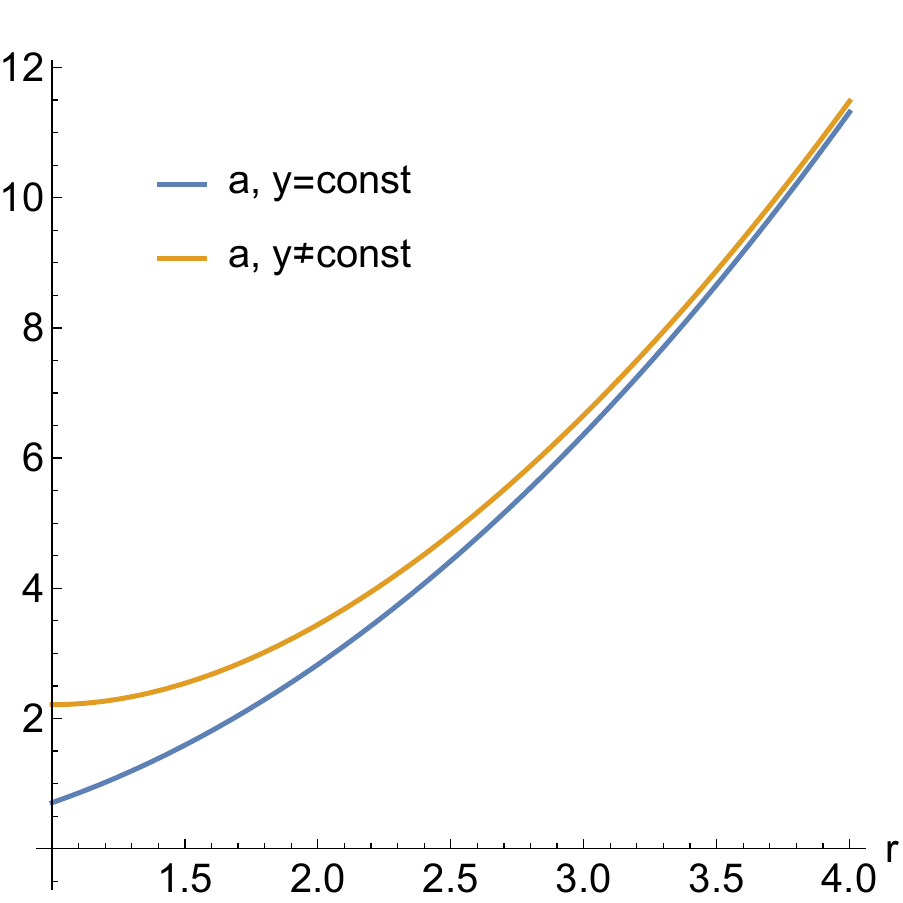}
\caption{\label{fig:sol3} Black hole solutions for $\kappa=0, C_1=0$ and $\beta=1$, $\Lambda=2$, $M=1$. On the left panel the metric function $h$ is depicted 
as a function of the radial coordinate for the solution~(\ref{LifshBH}) and for the solution~(\ref{hk0}), which asymptotes~(\ref{LifshBH}) for large $r$.
On the left panel the radial component of the vector field is shown as a function of $r$ for the same solutions.}
\end{figure}

\section{Conclusions}

In this paper we have undertaken a complete analysis of spherically symmetric and planar solutions of extended Proca vector tensor theory (\ref{action}). We found static black holes but also for the first time soliton solutions. 
We studied the general problem for (\ref{action}) explicitly reducing the system to two master equations, one algebraic and a non linear ODE, 
see Eqs.~(\ref{master1}) and (\ref{master2}). The solutions we found are generically of adS asymptotics in the presence of a Proca mass term $\mu$ while they are asymptotically flat in the absence of $\mu$ (and $\Lambda$) but, generic $\beta$, see Sec~\ref{sec:beta}. 
This is in contrast to the claim of \cite{Chagoya:2016aar} where it was found that only $\beta=1/4$ could sustain asymptotically flat black holes. It is important to note that it is due to the higher order vector galileon term  that solutions with regular asymptotics exist, in contrast to the case of pure Proca theory.
In the extended Proca theory, the Proca mass term plays the role of an effective negative cosmological constant.
An imaginary Proca mass term would permit de Sitter asymptotics. 

Furthermore, for the case of $\beta=1/4$ we found regular soliton solutions by putting the integration constant $M$ to zero, see Sec.~\ref{sec:special}. 
The massive vector nature of the solution regularises the spacetime metric and the solutions are particle like lumps of matter. 
One may need to use numerics to see if this property can be extended for $\beta\neq 1/4$ and $C_1\neq 0$.
Other solutions we found for $M=0$, aside from $\beta=1/4$, have a singularity which is however always hidden behind an event horizon. Therefore in these extended Proca theories solutions are more regular than in standard Einstein Maxwell (EM) theory. Indeed in GR with electric field we have a RN solution which is however singular for small mass compared to electric charge.
In other words, within this modified EM theory the vectorial mass term helps in regularizing spacetime solutions giving in certain cases gravitational particle like solitons.  
We also found Lifshitz spacetime solutions which have the characteristic to require only one vector field \cite{Cvetic:2014nta}, \cite{Liu:2014tra}. The Lifshitz coefficient is set by the coupling $\beta$. 

In EM theory electric and magnetic solutions are identical due to the electromagnetic duality.
Here, the theory we have studied, breaks electromagnetic duality and magnetic solutions should be found anew. 
Also it would be interesting to study the combination of the higher order curvature term with the Horndeski Maxwell term~\cite{Horndeski:1977kz}. 
These are some of the issues that may be worth pursuing in the near future.  

\section*{Acknowledgements}
EB and CC are grateful to the IMAFI institute in Talca-Chile for hospitality and financial support during the course of this work.  MH thanks the LabEx P2IO of Paris Saclay for financial support during the initial stages of this work.  EB and CC acknowledge support from the research program, Programme national de cosmologie et Galaxies of the CNRS/INSU, France and from the project DEFI InFIniTI 2017. EB was supported in part by Russian Foundation for Basic Research Grant No. RFBR 15-02-05038. MH is partially supported by grant 1130423 from FONDECYT and from CONICYT, Departamento de Relaciones Internacionales ÒPrograma Regional MATHAMSUD 13 MATH-05Ó. This project is also partially funded by Proyectos CONICYT- Research Council
UK DPI20140053. CC thanks Chiara Caprini  and Christos Tsagas for interesting discussions on cosmological magnetic fields and Lavinia Heisenberg for making some critical remarks in the introduction. 

\appendix
\section{Integration of the metric function for the case $\kappa=1$, $C_1=0$}
\label{app int}
Substituting the expressions (\ref{rofy}) and (\ref{aofy}) into (\ref{hofy}), we find,
Substituting the expressions (\ref{rofy}) and (\ref{aofy}) into (\ref{hofy}), we obtain for the metric function $h$,
\begin{equation}
h(r)=-\frac{2M}{r}
+\frac{\beta r_0 a_0^2}{2 r} I_1(y)
+\frac{r_0^3 a_0^2 \mu^2}{4\kappa\, r} I_2(y),
\end{equation}
where we introduced the notations,
\begin{equation}
	I_1 = \int \frac{dy}{y} \frac{(y+1-\gamma)^\frac{1+\gamma}{2\gamma}}{(y+1+\gamma)^\frac{1-\gamma}{2\gamma}},\;\;\;
	I_2 = \int \frac{dy}{y} (y+1-\gamma)^\frac{3\gamma-1}{2\gamma} (y+1+\gamma)^\frac{3\gamma+1}{2\gamma}.
\end{equation}
For the value $\beta=1/3$ one can find an explicit expression for the above integrals. Indeed in this case $\gamma=1/2$ and we find,
\begin{equation}
\label{int1}
	I_1(y) = 
	\int \frac{dy}{y} \frac{(y+\frac12)^\frac32}{(y+\frac32)^\frac12},\;\;\;
	I_2(y) = 
	\int \frac{dy}{y} \left(y+\frac12\right)^\frac12 \left(y+\frac32\right)^\frac52.
\end{equation}
To evaluate the above expressions we introduce a new variable $x$, such that
\begin{equation}
\label{xy}
x=\left(\frac{y+\frac12}{y+\frac32}\right)^{1/2}.
\end{equation}
With this change of the variable, the integrals in Eq.~(\ref{int1}) can be integrated explicitly.  Up to the constant of integration, they read,
\begin{equation}
\begin{split}
I_1(y) =& \int \frac{4x^4  dx}{(1-x^2)^2(3x^2-1)} = 
\int dx \left( \frac{1}{2(1-x)^2} + \frac{1}{2(1+x)^2} + \frac{1}{2(\sqrt{3}x-1)} - \frac{1}{2(\sqrt{3}x+1)}\right)\\
 &=  \frac{x}{1-x^2} + \frac{1}{2\sqrt{3}} \log\frac{\sqrt{3}x-1}{\sqrt{3}x+1}\\
& = \frac12 \sqrt{(1+2y)(3+2y)} + \frac{1}{2\sqrt{3}} \log \frac{3+4y-\sqrt{3(1+2y)(3+2y)}}{2y},
\end{split}
\end{equation}
and 
\begin{equation}
\begin{split}
I_2(y)= & \int \frac{4x^2  dx}{(1-x^2)^4(3x^2-1)} \\
& = \frac{7x}{4(1-x^2)}+\frac{x}{3(1-x^2)^3} + \frac{2x}{3(1-x^2)^2} +2\log\frac{1+x}{1-x} + \frac{9\sqrt{3}}{8}\log\frac{\sqrt{3}x-1}{\sqrt{3}x+1}\\
&= \frac{1}{12}\sqrt{(1+2y)(3+2y)}\left( 21+ 10y + 2y^2 \right) + \frac{9\sqrt{3}}{8} \log \frac{3+4y-\sqrt{3(1+2y)(3+2y)}}{2y}\\
& + 2\log\left(2 +2y +\sqrt{(1+2y)(3+2y)}\right).
\end{split}
\end{equation}
The asymptotic behaviour of the functions $I_1$ and $I_2$ is, 
\begin{equation}
\begin{split}
	I_1(y) \sim \log (y),\;\; I_2(y) \sim \frac{9\sqrt{3}}{8}\log(y) & \quad \text{for} \quad y\to 0,\\
	I_1(y) \sim y + \left(1+\frac{\log(2-\sqrt{3})}{2\sqrt{3}}\right),\;\; I_2(y) \sim \frac{y^3}3 & \quad \text{for} \quad y\to \infty.
\end{split}
\end{equation}


\end{document}